\documentclass[useAMS,usenatbib]{mn2e}
\usepackage{amsfonts}
\usepackage{amsmath}
\usepackage{amssymb}

\usepackage[T1]{fontenc}

\usepackage[dvips]{graphicx}
\usepackage{myaasmacros}
\usepackage{color}
\usepackage{multirow}

\def\bea{\begin{eqnarray}}
\def\ena{\end{eqnarray}}

\newcommand{\mr}[1]{\mathrm{#1}}

\title[The Fermi LAT view of the CWBs.]{The Fermi LAT view of the colliding wind binaries.}

\author[M. S. Pshirkov]{M. S. Pshirkov$^{1,2,3}$\thanks{E-mail: pshirkov@sai.msu.ru}\\
$^{1}$Sternberg Astronomical Institute, Lomonosov Moscow State University, Universitetsky prospekt 13, 119992, Moscow, Russia\\
$^{2}$Pushchino Radio Astronomy Observatory, 142290 Pushchino, Russia\\
$^{3}$Institute for Nuclear Research of the Russian Academy of Sciences, 117312, Moscow, Russia\\
}

\begin{document}

\date{}

\pubyear{2015}

\maketitle

\label{firstpage}

\begin{abstract}
Colliding wind binaries (CWBs) have been considered as a possible high energy  $\gamma$-ray sources for some time, however no system  other than $\eta$ Car has been detected. In the paper a sample of seven CWBs (WR 11, WR 70, WR 137, WR 140, WR 146, WR 147) which, by means of  theoretic modelling,   were deemed most promising candidates, was analyzed using almost 7 years of the Fermi-LAT data.
WR 11 ($\gamma^2$ Vel) was detected at 6.1$\sigma$ confidence level with a  photon flux in 0.1-100 GeV range  $(1.8\pm0.6)\times10^{-9}~\mr{ph~cm^{-2}~s^{-1}}$ and an energy flux  $(2.7\pm0.5)\times10^{-12}~~\mr{erg~cm^{-2}~s^{-1}}$. At the adopted distance $d=340$ pc this corresponds to  a luminosity $L=(3.7\pm0.7)\times10^{31}~\mr{erg~s^{-1}}$. This luminosity amounts to  $\sim6\times10^{-6}$  fraction of the  total wind kinetic power and $\sim1.6\times10^{-4}$  fraction of  the power injected into the wind-wind interaction region of this system. Upper limits were set on the high energy flux from the   WR 70 and  WR 140 systems. 

\end{abstract}

\begin{keywords}
gamma-rays:stars - stars:binaries - stars: Wolf-Rayet - stars:winds,outflows - stars:individual:WR11, WR70, WR125, WR137, WR140, WR146, WR147

\end{keywords}

%%%%%%%%%%%%%%%%%%%%%%%%%%%%%%%%%%%%%%%%%%%%%%%%%%%%%%%%%%%%%%%%%%%
\section{Introduction}
\label{sec:intro}
%%%%%%%%%%%%%%%%%%%%%%%%%%%%%%%%%%%%%%%%%%%%%%%%%%%%%%%%%%%%%%%%%%%
Hot massive stars of early types have very strong stellar winds. In the specific case of Wolf-Rayet(WR) stars the  mass loss rate could reach $10^{-4}~\mr{M_{\odot}~yr^{-1}}$ with terminal velocities $>1000~\mr{km~s^{-1}}$,  wind kinetic power of these stars could exceed $10^{37}~\mr{erg~s^{-1}}$. 

Massive binaries where both stars belong to early types and launch strong winds could be prominent sources of non-thermal radiation. Colliding supersonic winds produce strong shocks that could accelerate particles  to ultrarelativistic energies, harnessing up to several percent of the total kinetic wind power \citep{DeBecker2013}. The acceleration takes place in a region  where the energy density of the  photon and magnetic fields is very high. This leads to a  production of a strong non-thermal radiation via synchrotron and inverse Compton mechanisms \citep{Eichler1993,Dougherty2000,Benaglia2003,Bednarek2005,Reimer2006,Pittard2006,Reimer2009,Reitberger2014}. Another contribution to the high energy (HE) part of the spectrum  could arise due to the presence of accelerated hadrons that  could interact with wind material, producing pions. 
Non-thermal synchrotron radiation from several of  colliding wind binaries (CWBs) has already been discovered, in some cases interferometric observations reveal spatially extended region of the synchrotron emission \citep{Dougherty1996,Williams1997}.

No HE radiation from the CWBs has been observed besides truly exceptional $\eta$ Car system \citep{Tavani2009,Bednarek2011,Reitberger2015} and previous searches in the  HE  domain produced only upper limits for other systems \citep{Tatisheff2004,Werner2013}. 
Seven most promising  candidate systems were selected and analyzed using  2 years of the Fermi LAT data in \citep{Werner2013}.  In this paper we reanalyze this set using almost 7 years of data and the latest event reconstruction Pass 8. 
The rest of this paper is organized as follows: in section \ref{sec:data_methods} the data and the method used are introduced,  section \ref{sec:results} includes the results and their interpretation, the conclusions are presented in the section \ref{sec:summary}.

%%%%%%%%%%%%%%%%%%%%%%%%%%%%%%%%%%%%%%%%%%%%%%%%%%%%%%%%%%%%%%%%%%%
\section{Data and method}
\label{sec:data_methods}
%%%%%%%%%%%%%%%%%%%%%%%%%%%%%%%%%%%%%%%%%%%%%%%%%%%%%%%%%%%%%%%%%%%
%\paragraph*{The data.} 

A list of the most promising candidates for  detection  by the Fermi LAT was compiled in \citep{Werner2013}, see Table \ref{tab:sources}. In this paper these candidates were reanalyzed using  almost 7 years of new Pass 8 Fermi LAT data  from 2008 August 4 to 2015 July 1. All events in the energy range 0.1-100 GeV within 15$^{\circ}$ of the CWB positions  belonging to the ``SOURCE'' class 
  were selected and the recommended quality cut (zenith angle less than 90$^\circ$) was applied.  
After that, binned likelihood analysis was performed using  Fermi science tools version \textit{v10r0p5}\footnote{http://fermi.gsfc.nasa.gov/ssc/data/analysis/software/}.

Implemented source models included  the sources from the
3FGL catalogue \citep{3FGL}, the latest galactic interstellar emission
model \textit{gll\_iem\_v06.fits}, and the isotropic spectral template
\textit{iso\_P8R2\_SOURCE\_V6\_v06.txt}\footnote{http://fermi.gsfc.nasa.gov/ssc/data/access/lat/\\BackgroundModels.html}. Parameters of the sources inside the regions of interest (RoI) were allowed to change. Additional  $\gamma$-ray sources from the 3FGL catalogue between $15^{\circ}$ and $25^{\circ}$ from the RoI centers were included with fixed parameters. The candidate sources were initially modelled with a simple power-law spectral model.

% \,  in the subsequent analysis their parameters were held fixed.

\begin{table}
\caption{\label{tab:sources} List of the candidate sources} 
\begin{tabular}{|c|c|c|c|c|c|}
\hline
Name & $\alpha$,  h m s  &  $\delta$, d m s   &  $l,^\circ$  & $b,^\circ$ & Distance, kpc \\
\hline
WR 11 & 	08 09 32  & -47 20 12  &  262.80   & -07.69 &  0.34 \\

\hline
WR 70 & 	15 29 45     & -58 34 51   &  322.34    & -1.81 &  1.9 \\
\hline
WR 125 & 	19 28 16      & +19 33 21   &  54.44     & +1.06 &  3.1 \\
\hline
WR 137 & 20 14 32      & +36 39 40   &  74.33    &  +1.10   &  2.4 \\
\hline
WR 140 & 20 20 28       & +43 51 16   &  80.93     &  +4.18   &  1.7 \\
\hline
WR 146 & 20 35 47        & +41 22 45   &  80.56      &  +0.44   &  1.2 \\
\hline
WR 147 & 20 36 44        & +40 21 08   &  79.85       &   -0.31   &  0.65\\

\hline
\end{tabular}
\end{table}

\section{Results and discussion}
\label{sec:results}
\begin{figure}
\centering
% \begin{picture}(220,150)
% {\includegraphics[width=.9\columnwidth]{WR147.eps}}
% \end{picture}
\includegraphics[width=.75\columnwidth]{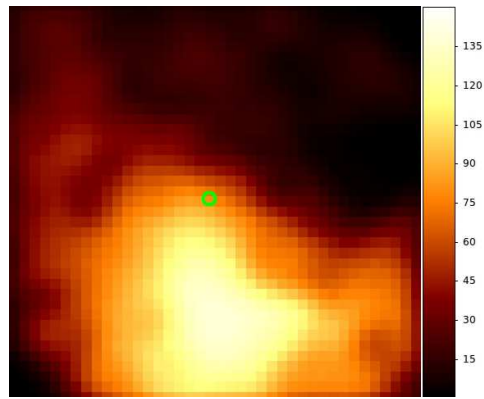}
\caption[width=1.0\columnwidth]{$TS$ map of 2x2 degree square centered at the WR 147 position. $TS$ excess is not well localized and the WR 147 position (indicated with the green circle) is away from the position of the $TS$ peak. }
\label{fig:tsmap_wr147}
\end{figure}

\begin{figure}
\centering
%\begin{picture}(220,150)
% \put(-10,0){\includegraphics[width=1.0\columnwidth]{WR11.eps}}
{\includegraphics[width=.75\columnwidth]{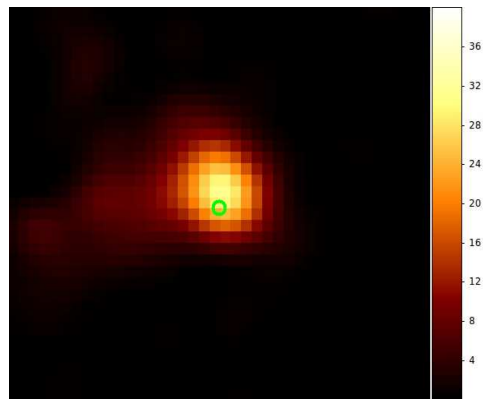}}
%\end{picture}
\caption[width=1.0\columnwidth]{$TS$ map of 2x2 degree square centered at the WR 11 position. $TS$ excess is  localized around the WR 11 position (indicated with the green circle). }
\label{fig:tsmap_wr11}
\end{figure}

%%%%%%%%%%%%%%%%%%%%%%%%%%%%%%%%%%%%%%%%%%%%%%%%%%%%%%%%%%%%%%%%%%%
The results are presented in the Table \ref{tab:ts}. It could be seen that three candidates: WR 11, WR 125, and WR 147  have significance exceeding threshold $TS=25$ and the significance for the WR 37 system is hovering around this threshold ($TS=23.7$).  However, these sources reside close to the galactic plane where the $\gamma$-ray background is especially strong. In case of its imperfect modelling the likelihood analysis could possibly produce unrealistic results. In order to check the validity of  detections,  $TS$ maps of the immediate neighborhoods were calculated using the \emph{gttsmap} tool. Unfortunately, in all cases except the WR 11,  there were several  hotspots in $<0.5^{\circ}$ radius with  the significance larger than that of the ``sources'' (see for illustration WR 147,  Fig. \ref{fig:tsmap_wr147}). That is not the case for WR 11, where there is a clear maximum of $TS$ residing very close to the CWB (less than 0.05$^{\circ}$) (see Fig. \ref{fig:tsmap_wr11}). Thus it can be concluded that the source spatially coinciding with this system is genuine.

\begin{table}
\centering
\caption{\label{tab:ts} $TS$ of the candidate sources (simple power-law model).} 
\begin{tabular}{|c|c|}
\hline
Name & TS \\
\hline
WR 11 & 37.7 \\

\hline
WR 70 & 1.2 \\
\hline
WR 125 & 41.0 \\
\hline
WR 137 & 23.3 \\
\hline
WR 140 & 0.1 \\
\hline
WR 146 & 15.0 \\
\hline
WR 147 & 54.9 \\

\hline
\end{tabular}
\end{table}

\begin{figure}
\raggedleft
% \put(-10,0){\includegraphics[width=1.0\columnwidth]{WR11.eps}}http://www.rbc.ru/
{\includegraphics[angle=270,width=.9\columnwidth]{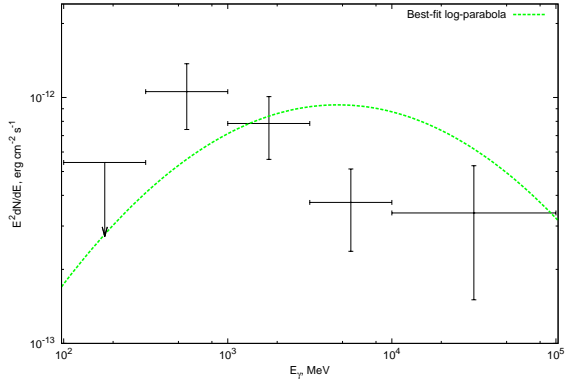}}

\caption[width=.9\columnwidth]{The spectrum of the WR11 system. It could be seen that the 'best-fit' curve fits rather poorly due to the presence of an additional hard  tail at the energies $E>10$ GeV. This spectrum strongly resembles the spectrum of $\eta$ Car during its periastron passage \citep{Reitberger2015}. The upper limits indicated with arrows correspond to 95\% CL.}
\label{fig:spectrum}
\end{figure}

Results of the fit with a simple power-law model for this source, from now on dubbed  WR 11, gives a $TS=37.7$ and a spectral index $\Gamma = 2.16\pm0.2$. However, the spectrum of the source is considerably curved ( Fig. \ref{fig:spectrum}), and the fits with more elaborated spectral models fare a bit better: $TS=41.5$ for a log-parabola model and $TS=44.3$ for a  broken power-law model. Nevertheless, they all  fail to  fully reproduce  the spectral shape: there is a hard tail at the energies larger than 10 GeV. Overall spectral shape is very close to the one of $\eta$ Car during its periastron passage \citep{Reitberger2015}. 

The WR11 ($\gamma^2$ Vel, WC8+O7.5) at a distance $d=340$ pc is the closest to us Wolf-Rayet binary. The  components have small separation ($0.8-1.6$~au), and  the orbital period is short, $P=78.53$ days \citep{wr11}.  The orbital eccentricity is non-negligible, $e=0.33$.
 The relevant physical parameters are presented in the Table \ref{tab:WR11}. Total kinetic wind power $L_\mr{w}=5.8\times10^{36}~\mr{erg~s^{-1}}$ is dominated by the contribution from the Wolf-Rayet component. A fraction of the total kinetic power that is dissipated in the wind collision zone -- $L_{\mr{cwz}}$ --   could be estimated from  a purely geometrical reasoning and it is equal to the wind momentum ratio $\eta$  \citep{DeBecker2013}:
 \begin{equation}
  \eta=\frac{\dot{M}_{\mr{O}}v_{\mr{O}}^{\infty}}{\dot{M}_{\mr{WR}}v_{\mr{WR}}^{\infty}}=0.04,
  \label{eq:eta}
 \end{equation}
 \begin{equation}
  L_{\mr{cwz}}=\eta L_{\mr{w}}=2.3\times10^{35}~\mr{erg~s^{-1}}.
  \label{eq:Lcwz}
 \end{equation}
The  X-ray emission from the colliding wind shock  with an unabsorbed luminosity reaching $10^{33}~\mr{erg~s^{-1}}$  (0.4-10 keV energy range) was also detected \citep{Schild2004}. WR11 is a bright radio source  at the cm wavelengths (26.5 mJy at 4.8 GHz, $L_{\mr{rad}}=1.8\times10^{28}~\mr{erg~s^{-1}}$), but the bulk of the observed emission has a thermal origin \citep{Leitherer1997}.  The non-thermal signal from the shock would be strongly absorbed in the dense plasma of the stellar winds, eventually contributing up to a half of the total luminosity at 4.8 GHz ($L_{\mr{non-thermal}}=8.3\times10^{27}~\mr{erg~s^{-1}}$)\citep{Chapman1999}.
%   The WC8  star has a mass $M_{WR}=9.0~\mr{M_{\odot}}$ and  a mass-loss rate %$\dot{M}_{WR}=8\times10^{-6}~\mr{M_{\odot}~yr^{-1}}$, its  more massive ($M_{O}=29~\mr{M_{\odot}}$) companion has considerably smaller rate $\dot{M}_O=1.8\times10^{-7}~\mr{M_{\odot}~yr^{-1}}$.  or the WR component us estimated ti  
\begin{table}
\centering
\caption{\label{tab:WR11} Adopted physical parameters of the WR11 system. } 
\begin{tabular}{c|c|c|c|}
\hline
Parameter&unit&WC8& O7.5\\
\hline
Mass, $M$ &$\mr{M_{\odot}}$ & 9.0 & 29.0 \\
\hline
Mass-loss rate, $\dot{M}$&$10^{-7}~\mr{M_{\odot}~yr^{-1}}$ & 80 & 1.8 (1) \\
\hline
Terminal wind velocity, $v^{\infty}$& $\mr{km~s^{-1}}$ & 1450 & 2500 (1) \\
\hline
Luminosity, $L$ &$10^5~\mr{L_{\odot}}$&1.7 & 2.8 \\
\hline
\end{tabular}
\textbf{References}: If not otherwise specified, all values  are taken from \citep{wr11}; (1) \citep{DeMarco1999}.
\end{table}

The detected $\gamma$-ray source is  rather weak: the flux in the HE range (0.1 -100 GeV) from the WR 11 is $(1.8\pm0.6)\times10^{-9}~\mr{ph~cm^{-2}~s^{-1}}$, and the energy flux $(2.7\pm0.5)\times10^{-12}~~\mr{erg~cm^{-2}~s^{-1}}$. At the adopted distance $d=340$ pc this gives  a  luminosity $L=(3.7\pm0.7)\times10^{31}~\mr{erg~s^{-1}}$ or $\sim6\times10^{-6}$ fraction of total kinetic wind power $L_W=5.8\times10^{36}~\mr{erg~s^{-1}}$ and $\sim1.6\times10^{-4}$ fraction of the wind power  that flows into the wind collision zone.

The mechanism of the emission is uncertain, though the latest simulations imply that the hadronic processes dominate when the binary separation is small, like in the case of WR 11 ($\sim10^{13}$~cm) \citep{Reitberger2014}. On the other hand these simulations predict almost flat spectrum around $E=100$ MeV for the hadronic mechanism and that could be in mild tension with the observations (see Fig. \ref{fig:spectrum}). Future observations, including  ones  in the extended energy range ($<100$ MeV and $>100$ GeV), will allow to clarify this issue.

If the emission is leptonic in origin then it is possible to estimate the magnetic field strength in the collision wind zone: the  ratio of the magnetic field energy density $\epsilon_{B}$ to the  seed photon field density $\epsilon_{\mr{ph}}$ is equal to the  ratio of  the $\gamma$-ray luminosity to the non-thermal radio luminosity \citep{Tatisheff2004}:
$${\epsilon_B}/{\epsilon_{\mr{ph}}}={L_{\mr{non-thermal}}}/{L_{\gamma}}\equiv K=2.2\times10^{-4},$$

 \begin{equation}
  B=\sqrt{\frac{2KL_{\mr{O}}}{r_{\mr{O}}^2c}}=\sqrt{\frac{2KL_{\mr{O}}}{\eta r^2c}}=1.7~\mr{G},
  \label{eq:mf}
 \end{equation}

where $L_{\mr{O}}=1.1\times10^{39}~\mr{erg~s^{-1}}$ is the O-type star luminosity, $r_{\mr{O}}=\sqrt{\eta}r$ is the distance from the  star to the collision zone which is  a fraction of the total separation $r=0.8~\mr{au}$. Magnetic field of this magnitude  could be expected when the collision zone is located at such a small distance from the O-type star with a surface field of $\sim100$ G strength \citep{Eichler1993,Tatisheff2004}. It is worth noting that even if the HE radiation is leptonic in origin it is produced by much more energetic population than the one producing the synchrotron emission: $\gamma\sim10^{4}$ rather than several tens. 

Finally, search for the periodicity corresponding to the binary period $P=78.53$ days was performed. None was found. Low statistics with  the total number of the photons from the source $<10^3$ probably precluded any observations of the periodical modulations.

The WR 11 system due to its proximity and relatively high galactic latitude remains the only detected system, only  upper limits on their flux could be calculated for the rest of the CWBs in the list. It could be meaningless in the cases of WR 125, WR 137,  and WR 147  with their spurious high $TS$s, so the limits were set only for WR 70 and WR 140 (see Tab. \ref{tab:UL}). WR 146 is a borderline case: with its  $TS\sim15$ and  complicated neighbourhood the calculated ULs could also be difficult to interpret.

\begin{table}
\centering
\caption{\label{tab:UL} Upper limits  on the flux (95\% CL)  in the 0.1-100 GeV energy range}
\begin{tabular}{|c|c|}
\hline
Name & $F_{100},~10^{-9}\mr{ph~ cm^{-2}~s^{-1}}$ \\ 
\hline
WR 70 & 2.6 \\

\hline
WR 140 & 1.1 \\
\hline

\end{tabular}
\end{table}

%%%%%%%%%%%%%%%%%%%%%%%%%%%%%%%%%%%%%%%%%%%%%%%%%%%%%%%%%%%%%%%%%%%
\section{Summary}\label{sec:summary}
%%%%%%%%%%%%%%%%%%%%%%%%%%%%%%%%%%%%%%%%%%%%%%%%%%%%%%%%%%%%%%%%%%%

A search for  HE emission from  seven  potential  candidates was conducted. HE emission from the nearest system WR 11 ($\gamma^2$ Vel) was detected at 6.1$\sigma$
confidence level ($TS=37.7$) with a simple power-law model (spectral index $\Gamma=2.16\pm0.2$). The spectrum of the source is curved with a broad maximum around $\sim$1 GeV and shows a  hardening at energies above 10 GeV. The photon flux in 0.1-100 GeV range  is  $(1.8\pm0.6)\times10^{-9}~\mr{ph~cm^{-2}~s^{-1}}$, the energy flux is   $(2.7\pm0.5)\times10^{-12}~~\mr{erg~cm^{-2}~s^{-1}}$. At the adopted distance $d=340$ pc that corresponds to luminosity $L=(3.7\pm0.7)\times10^{31}~\mr{erg~s^{-1}}$, that is   $\sim6\times10^{-6}$  fraction of total wind kinetic power and $\sim1.6\times10^{-4}$  fraction of  power injected into the wind interaction region of this system.  
Upper limits were set on the HE luminosity of WR 70 and   WR 140 systems.
The detection of WR 11 by the Fermi LAT endorses colliding wind binaries as a new separate class of high energy sources.
%%%%%%%%%%%%%%%%%%%%%%%%%%%%%%%%%%%%%%%%%%%%%%%%%%%%%%%%%%%%%%%%%%%
\section*{Acknowledgements}
%%%%%%%%%%%%%%%%%%%%%%%%%%%%%%%%%%%%%%%%%%%%%%%%%%%%%%%
The author would like  to thank the anonymous referee for the valuable comments that helped to improve the quality of the paper.
This research was supported by RSF grant No. 14-12-00146. The author acknowledges the fellowship of the Dynasty Foundation. The numerical part of the
work was done at the cluster of the Theoretical Division of INR RAS.
This research has made use of NASA's Astrophysics Data System.

%%%%%%%%%%%%%%%%%%%%%%%%%%%%%%%%%%%%%%%%%%%%%%%%%%%%%%%%%%%%%%%%%%%
%\section*{Appendix}
%%%%%%%%%%%%%%%%%%%%%%%%%%%%%%%%%%%%%%%%%%%%%%%%%%%%%%%%%%%%%%%%%%%

%\newpage

\bibliographystyle{mn2e}
\bibliography{cwb}

\begin{thebibliography}{}

\bibitem[\protect\citeauthoryear{{Bednarek}}{{Bednarek}}{2005}]{Bednarek2005}
{Bednarek} W.,  2005, \mnras, 363, L46

\bibitem[\protect\citeauthoryear{{Bednarek} \& {Pabich}}{{Bednarek} \&
  {Pabich}}{2011}]{Bednarek2011}
{Bednarek} W.,  {Pabich} J.,  2011, \aap, 530, A49

\bibitem[\protect\citeauthoryear{{Benaglia} \& {Romero}}{{Benaglia} \&
  {Romero}}{2003}]{Benaglia2003}
{Benaglia} P.,  {Romero} G.~E.,  2003, \aap, 399, 1121

\bibitem[\protect\citeauthoryear{{Chapman}, {Leitherer}, {Koribalski}, {Bouter}
  \& {Storey}}{{Chapman} et~al.}{1999}]{Chapman1999}
{Chapman} J.~M.,  {Leitherer} C.,  {Koribalski} B.,  {Bouter} R.,    {Storey}
  M.,  1999, \apj, 518, 890

\bibitem[\protect\citeauthoryear{{De Becker} \& {Raucq}}{{De Becker} \&
  {Raucq}}{2013}]{DeBecker2013}
{De Becker} M.,  {Raucq} F.,  2013, \aap, 558, A28

\bibitem[\protect\citeauthoryear{{De Marco} \& {Schmutz}}{{De Marco} \&
  {Schmutz}}{1999}]{DeMarco1999}
{De Marco} O.,  {Schmutz} W.,  1999, \aap, 345, 163

\bibitem[\protect\citeauthoryear{{Dougherty} \& {Williams}}{{Dougherty} \&
  {Williams}}{2000}]{Dougherty2000}
{Dougherty} S.~M.,  {Williams} P.~M.,  2000, \mnras, 319, 1005

\bibitem[\protect\citeauthoryear{{Dougherty}, {Williams}, {van der Hucht},
  {Bode} \& {Davis}}{{Dougherty} et~al.}{1996}]{Dougherty1996}
{Dougherty} S.~M.,  {Williams} P.~M.,  {van der Hucht} K.~A.,  {Bode} M.~F.,
  {Davis} R.~J.,  1996, \mnras, 280, 963

\bibitem[\protect\citeauthoryear{{Eichler} \& {Usov}}{{Eichler} \&
  {Usov}}{1993}]{Eichler1993}
{Eichler} D.,  {Usov} V.,  1993, \apj, 402, 271

\bibitem[\protect\citeauthoryear{{Fermi-LAT Collaboration}}{{Fermi-LAT
  Collaboration}}{2015}]{3FGL}
{Fermi-LAT Collaboration} 2015, \apjs, 218, 23

\bibitem[\protect\citeauthoryear{{Leitherer}, {Chapman} \&
  {Koribalski}}{{Leitherer} et~al.}{1997}]{Leitherer1997}
{Leitherer} C.,  {Chapman} J.~M.,    {Koribalski} B.,  1997, \apj, 481, 898

\bibitem[\protect\citeauthoryear{{North}, {Tuthill}, {Tango} \&
  {Davis}}{{North} et~al.}{2007}]{wr11}
{North} J.~R.,  {Tuthill} P.~G.,  {Tango} W.~J.,    {Davis} J.,  2007, \mnras,
  377, 415

\bibitem[\protect\citeauthoryear{{Pittard} \& {Dougherty}}{{Pittard} \&
  {Dougherty}}{2006}]{Pittard2006}
{Pittard} J.~M.,  {Dougherty} S.~M.,  2006, \mnras, 372, 801

\bibitem[\protect\citeauthoryear{{Reimer}, {Pohl} \& {Reimer}}{{Reimer}
  et~al.}{2006}]{Reimer2006}
{Reimer} A.,  {Pohl} M.,    {Reimer} O.,  2006, \apj, 644, 1118

\bibitem[\protect\citeauthoryear{{Reimer} \& {Reimer}}{{Reimer} \&
  {Reimer}}{2009}]{Reimer2009}
{Reimer} A.,  {Reimer} O.,  2009, \apj, 694, 1139

\bibitem[\protect\citeauthoryear{{Reitberger}, {Kissmann}, {Reimer} \&
  {Reimer}}{{Reitberger} et~al.}{2014}]{Reitberger2014}
{Reitberger} K.,  {Kissmann} R.,  {Reimer} A.,    {Reimer} O.,  2014, \apj,
  789, 87

\bibitem[\protect\citeauthoryear{{Reitberger}, {Reimer}, {Reimer} \&
  {Takahashi}}{{Reitberger} et~al.}{2015}]{Reitberger2015}
{Reitberger} K.,  {Reimer} A.,  {Reimer} O.,    {Takahashi} H.,  2015, \aap,
  577, A100

\bibitem[\protect\citeauthoryear{{Schild}, {G{\"u}del}, {Mewe}, {Schmutz},
  {Raassen}, {Audard}, {Dumm}, {van der Hucht}, {Leutenegger} \&
  {Skinner}}{{Schild} et~al.}{2004}]{Schild2004}
{Schild} H.,  {G{\"u}del} M.,  {Mewe} R.,  {Schmutz} W.,  {Raassen} A.~J.~J.,
  {Audard} M.,  {Dumm} T.,  {van der Hucht} K.~A.,  {Leutenegger} M.~A.,
  {Skinner} S.~L.,  2004, \aap, 422, 177

\bibitem[\protect\citeauthoryear{{Tatischeff}, {Terrier} \&
  {Lebrun}}{{Tatischeff} et~al.}{2004}]{Tatisheff2004}
{Tatischeff} V.,  {Terrier} R.,    {Lebrun} F.,  2004, in {Schoenfelder} V.,
  {Lichti} G.,   {Winkler} C.,  eds, 5th INTEGRAL Workshop on the INTEGRAL
  Universe Vol.~552 of ESA Special Publication, {High-Energy Emission from the
  Stellar Wind Collision in {$\gamma$}$^{2}$ Velorum}.
p.~409

\bibitem[\protect\citeauthoryear{Tavani et~al.,}{Tavani
  et~al.}{2009}]{Tavani2009}
Tavani M.,  et~al., 2009, Astrophys. J., 698, L142

\bibitem[\protect\citeauthoryear{{Werner}, {Reimer}, {Reimer} \&
  {Egberts}}{{Werner} et~al.}{2013}]{Werner2013}
{Werner} M.,  {Reimer} O.,  {Reimer} A.,    {Egberts} K.,  2013, \aap, 555,
  A102

\bibitem[\protect\citeauthoryear{{Williams}, {Dougherty}, {Davis}, {van der
  Hucht}, {Bode} \& {Setia Gunawan}}{{Williams} et~al.}{1997}]{Williams1997}
{Williams} P.~M.,  {Dougherty} S.~M.,  {Davis} R.~J.,  {van der Hucht} K.~A.,
  {Bode} M.~F.,    {Setia Gunawan} D.~Y.~A.,  1997, \mnras, 289, 10

\end{thebibliography}

%%%%%%%%%%%%%%%%%%%%%%%%%%%%%%%%%%%%%%%%%%%%%%%%%%%%%%%%%%%%%%%%%%%%%%%%%%%%%%%%%%%%%%%
%%%%%%%%%%%%%%%%%%%%%%%%%%%%%%%%%%%%%%%%%%%%%%%%%%%%%%%%%%%%%%%%%%%%%%%%%%%%%%%%%%%%%%%%

\end{document}